\documentclass[iop]{emulateapj}

\usepackage{graphicx}
\usepackage{natbib}
\usepackage{amssymb,txfonts}
\usepackage{multirow}
\usepackage{array}
\usepackage{sidecap}
\usepackage{hyperref}
\usepackage{epstopdf}
\usepackage{footnote}
\usepackage{tabularx}
\usepackage{booktabs}
\usepackage{longtable}
\usepackage{tabu}

\newcommand{\kms}{km~s$^{-1}$}

\newcommand{\YOHKOH}{\textit{Yohkoh}}
\newcommand{\Hinode}{\textit{Hinode}}

\newcommand{\sdo}{\textit{SDO}}

\shortauthors{Panesar et al.}

\begin{document}
 \title{Magnetic Flux Cancellation as the Trigger of Solar-Quiet Region Coronal Jets}

\author{Navdeep K. Panesar\altaffilmark{1}, Alphonse C. Sterling\altaffilmark{1}, Ronald L. Moore\altaffilmark{1,2}, Prithi Chakrapani\altaffilmark{3}}

\affil{{$^1$}Heliophysics and Planetary Science Office, ZP13, Marshall Space Flight Center, Huntsville, AL 35812, USA}

\affil{{$^2$}Center for Space Plasma and Aeronomic Research (CSPAR), UAH, Huntsville, AL 35805, USA}

\affil{{$^3$}Hunter College High School, New York, NY, USA}

\email{navdeep.k.panesar@nasa.gov}


\begin{abstract}
	We report observations of ten random on-disk solar quiet region coronal jets found in high resolution Extreme Ultraviolet (EUV) images from the 
	\textit{Solar Dynamics Observatory (SDO)}/Atmospheric Imaging Assembly (AIA) and having good coverage in magnetograms from the \sdo/Helioseismic and Magnetic Imager (HMI). Recent studies show that coronal jets are driven by the eruption of a  small-scale filament (called a \textit{minifilament}). However the trigger of these eruptions is still unknown. 
	In the present study we address the question: what leads to the jet-driving minifilament eruptions? The EUV observations show that there is a cool-transition-region-plasma minifilament present prior to each jet event and the minifilament eruption drives the jet. By examining  pre-jet evolutionary changes in the line-of-sight photospheric magnetic field we observe that each pre-jet  minifilament resides over the neutral line between majority-polarity and minority-polarity patches of magnetic flux. In each of the ten cases, the opposite-polarity patches approach and merge with each other (flux reduction between 21 and 57\%). After several hours, continuous flux cancellation at the neutral line apparently destabilizes the field holding the cool-plasma minifilament to erupt and undergo internal reconnection, and external reconnection with the surrounding-coronal field. The external reconnection opens the minifilament field allowing the minifilament material to escape outwards, forming part of the jet spire. Thus we found that each of the ten jets resulted from eruption of a minifilament following flux cancellation at the neutral line under the minifilament. These observations establish that magnetic flux cancellation is usually the trigger of quiet region coronal jet eruptions.
	
\end{abstract}

\section{INTRODUCTION}
Solar coronal jets are frequent magnetically channeled narrow eruptions observed in the solar corona \citep{raouafi16}. They are relatively 
short lived and transient features, occurring in various solar environments including quiet regions \citep{wang98,hong11}, coronal holes \citep{shimojo98,cirtain07,adams14} and active regions 
\citep{shibata92,innes11,panesar16,sterling16}. They have been often observed in 
 extreme ultraviolet (EUV) \citep{ywang98} and X-ray \citep{shibata92,canfield96,alexander99} emission. 
X-ray jets are well-studied and imaged by \YOHKOH\ and \Hinode. Most X-ray jets
have a lifetime of about 10 minutes, velocities of around 200 \kms\ and lengths of $\sim$ 5 $\times$ 10${^4}$ \kms\ \citep{shimojo96,savcheva07}. 
It has been observed that X-ray jets show a bright point (also known as jet bright point, JBP) at an edge of the base during the eruption \citep{shibata92}. Properties (e.g. velocities, lifetimes and a JBP) similar to X-ray jet properties have been seen in EUV coronal jets \citep{raouafi08,nistico09,pucci13,schmieder13}.

 A possible driving mechanism for jet eruptions is explosive magnetic reconnection.  However, the triggering and driving mechanisms are still not fully understood. Some workers have suggested that flux emergence may lead to the jet eruption \citep[e.g.][]{shibata92,shibata07,moreno08}. A few on-disk studies showed evidence of flux cancellation leading to the jet eruption \citep{hong11,huang12,shen12,adams14,young14b,young14a}, 
but until now there have been no systematic observational studies of the magnetic origin of jet eruptions. \cite{sterling15} analyzed 20 random coronal jets in polar coronal holes using soft X-ray and EUV images and found that those X-ray jets are driven by small-scale filament eruptions. Their study included only near-limb events that therefore lacked adequate magnetic field data, and so the question of what leads to these \textit{minifilament} eruptions remained open.

In this Letter, we investigate the eruption mechanism of ten random jets in on-disk quiet regions. We study the photospheric magnetic field evolution leading to the jet eruptions. That is, we track the evolution of the photospheric magnetic flux that leads to the minifilament eruption in jets. 
A key question we address in this Letter is, what causes the  jet eruptions: magnetic flux cancellation or flux emergence? By studying ten random jets, we find that flux cancellation is the cause of most quiet-region jet eruptions.


\begin{table*}
	\begin{center}
	
		\caption{Measured parameters for the observed quiet region jets \label{tab:list}}
	\small
		\begin{tabular}{c*{10}{c}}
			\noalign{\smallskip}\tableline\tableline \noalign{\smallskip}
			Event & Date & Time\tablenotemark{a} &  Location\tablenotemark{b} &  Jet Speed\tablenotemark{c} &  Jet Dur.\tablenotemark{d} &  Jet-base\tablenotemark{e} & Minifil. length\tablenotemark{f} & $\Phi$ values\tablenotemark{g} & \% of $\Phi$\tablenotemark{h}\\
			
			No.   &    &  (UT)   & x,y (arcsec) &(\kms)& min. & Width (km) & ($\pm$1700 km) &  10$^{19}$ Mx & reduction  \\
			
			\noalign{\smallskip}\hline \noalign{\smallskip}
			J1 & 2012 Mar 22  & 04:46  & -470,-100 &    100$\pm$30 & 15$\pm$5 & 10500$\pm$500 & 9800 &  1.6 & 52 $\pm$ 5.8 \\   [1ex]
			J2 & 2012 Jul 01  & 08:32  &  -44, 285    & 100$\pm$10   & 10$\pm$2 &27000$\pm$500 & 25000 &  4.0 &  18 $\pm$ 6.8      \\   [1ex]
			J3 & 2012 Jul 07  & 21:31  & -192,-180 &   120$\pm$15  & 14$\pm$3 &16500$\pm$400 & 10500 &  --\tablenotemark{i} & --  \\   [1ex]
			J4 & 2012 Aug 05\tablenotemark{j}  & 02:20  & -485, 190 &   140$\pm$35  & 10$\pm$3 &22000$\pm$1000 & 31000 &  1.5 &  21 $\pm$ 6.0  \\   [1ex]
			J5 & 2012 Aug 10  & 23:03  & -168,-443 &   125$\pm$15  & 15$\pm$2 &16000$\pm$400 & 10000 &  0.9 &  57 $\pm$ 5.4  \\  [1ex]
			J6 & 2012 Sept 20  & 22:56  & -158,-486    &  35$\pm$5  & 9$\pm$2 &20000$\pm$500 &  36000  &  2.0 &  23 $\pm$ 4.6      \\  [1ex]
			J7 & 2012 Sept 21  & 03:33  & -115, -485 &    135$\pm$30 & 12$\pm$1  &17500$\pm$500 & 15000  &  1.0 &  36 $\pm$ 7.2  \\   [1ex]
			J8 & 2012 Sept 22  & 01:25  & -338, 103 &    110$\pm$45  & 11$\pm$1  &13000$\pm$600 & 5700 &  0.9 &  50 $\pm$ 5.1  \\ [1ex]
			J9 & 2012 Nov 13  & 04:21  & -28,-307  &   55$\pm$5  & 9$\pm$3 &18000$\pm$1000 & 25000 &  1.7 &  34 $\pm$ 3.2  \\ [1ex]
			J10 & 2012 Dec 13  & 10:36  & 26, 50  &   65$\pm$20  & 10$\pm$2  &9500$\pm$500 & 12500 &  1.2 &  38 $\pm$ 5.0 \\ [1ex]
			\noalign{\smallskip}\tableline\tableline \noalign{\smallskip}
			
		\end{tabular}
		\footnotetext{Time of JBP approximate peak brightening in AIA 94 \AA\ images.}
		\footnotetext{Approximate location of the jet region on the solar disk.} 
		\footnotetext{Plane-of-sky speed along the jet spire (observed in AIA 171 \AA\ emission images), soon after jet started to erupt outwards. Speeds and uncertainties are estimated from the time-distance plots.  }
		\footnotetext{Duration of jet spire visibility in 171 \AA. }
		\footnotetext{Mean cross-sectional width of the jet-base region measured at the time of peak base brightening. }
		\footnotetext{Integrated length (along the curvilinear path) of the minifilament before eruption onset. }
		\footnotetext{ Average flux ($\Phi$) values of the minority flux clumps 5-6 hours before eruption.}
		\footnotetext{ Flux  change between 5-6 hours before eruption and 0-1 hours after eruption. }
		\footnotetext{ Flux patches are not isolated enough for a reliable measurement. }
		\footnotetext{There are two eruptions in the jet region, the first at 02:00 and the second at 02:20; J4 is the second eruption. }
		
	\end{center}
\end{table*}

\begin{figure*}
	\centering
	\includegraphics[width=\linewidth]{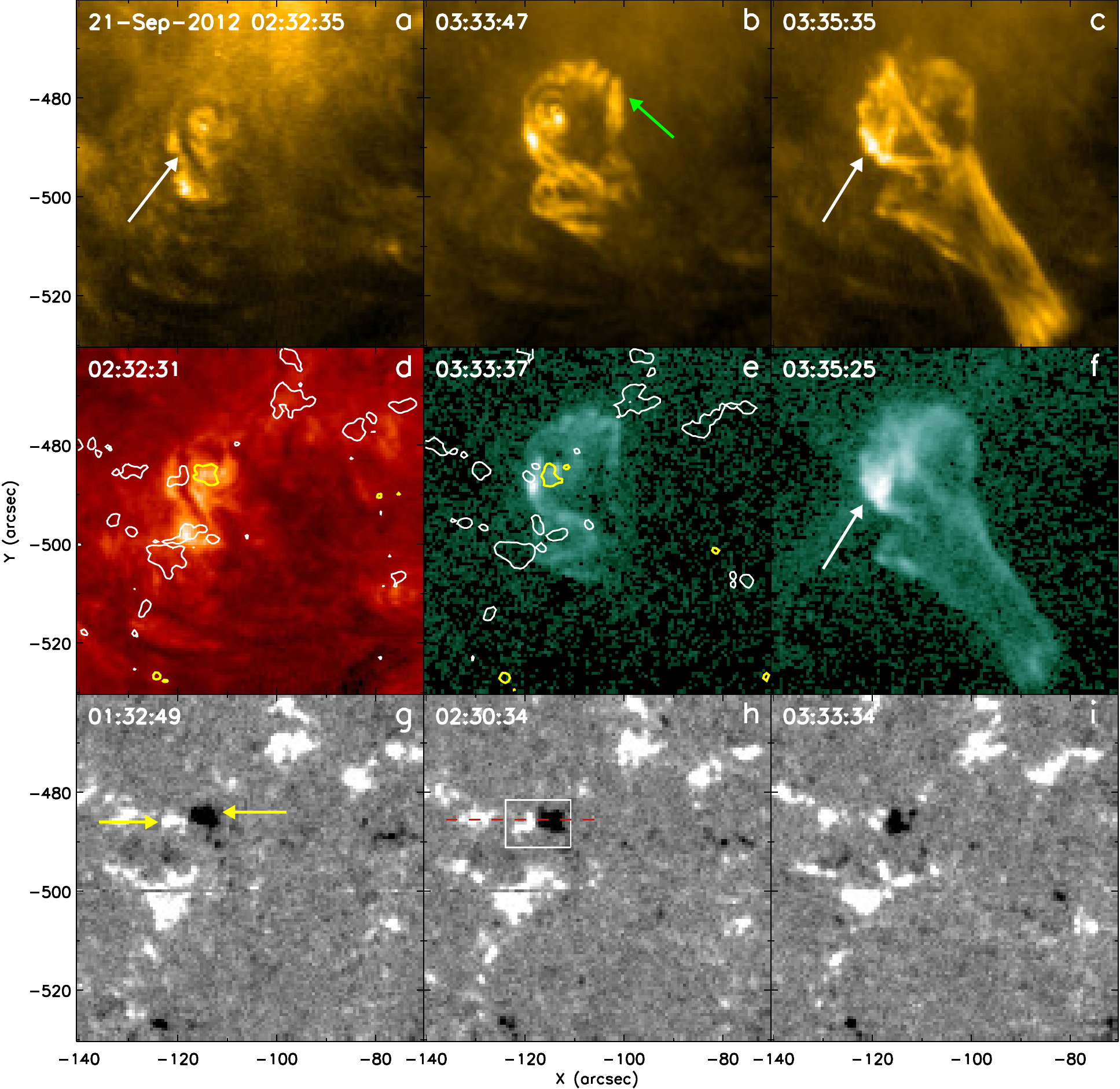}
	\caption{Quiet region jet observed on 21-Sept-2012: (a-c) show 171 \AA, (d) 304 \AA\ and (e,f) 94 \AA\ AIA intensity images of the jet J7 of Table \ref{tab:list}; (g-i) show HMI magnetograms of the same region. In (a), the arrow shows the minifilament. In (b), the green arrow points to brightening from external reconnection. In (c) and (f), the arrows point to the JBP. The yellow arrows in (g) point to converging positive (white) and negative (black) flux clumps; the pre-jet minifilament resides along the neutral line between these two clumps. The  boxed region in (h) shows the area measured for the magnetic flux time plot shown in Figure \ref{fig3}a; the red dashed line shows the east-west cut for the Figure \ref{fig3}b time-distance image. HMI contours (level $\pm$ 50 G) of (h) and (i) are overlaid on panels (d) and (e), respectively, where white and yellow respectively represent positive and negative
		polarities. 
		Animations of this Figure ((a-c) and (g-i)) are available.} \label{fig1}
\end{figure*} 

\vspace{0.3cm}
\section{INSTRUMENTATION AND DATA}\label{data}

 \textit{Solar Dynamics Observatory (SDO)}/Atmospheric Imaging Assembly (AIA) gives full-Sun images with high spatial resolution (0\arcsec.6 pixel$^{-1}$, $\sim$ 430 km) and high temporal cadence (12 s) in seven EUV wavelength bands \citep{lem12}. For the present study, we used multi-channel (304 \AA, 171 \AA, and 94 \AA) EUV images from \sdo/AIA to view cool-transition-region structures (minifilaments) and coronal-temperature jet structures (e.g. JBP). We primarily used 171 \AA\ images because we found pre-jet minifilaments to be best seen in this channel. 

We use line-of-sight magnetograms from the \sdo/Helioseismic and Magnetic Imager (HMI; \citealt{schou12}) with  high spatial resolution of 0\arcsec.5 pixel$^{-1}$ and temporal cadence of 45 s \citep{scherrer12} to examine the photospheric magnetic field of the jet region. With these magnetograms we follow the pre-jet evolution of the photospheric magnetic field of the jet-base region, while examining nearly-concurrent EUV images of coronal emission. 

We downloaded \sdo/AIA and \sdo/HMI data from the JSOC cutout service\footnote{http://jsoc.stanford.edu/ajax/exportdata.html}, and removed
  solar rotation by derotating all of the AIA and HMI images to a particular time. AIA and HMI data sets were co-aligned by using SolarSoft routines, and we over plotted the HMI contours of active-regions magnetic flux on to the AIA images to verify the alignment.
We created movies at two lower temporal cadences (1- and 5-minute for AIA and HMI, respectively) to study the jet dynamics and the jet-region magnetic evolution.

\vspace{0.3cm}
\section{RESULTS}\label{result}
\subsection{\textit{Overview}}
We examine the structure and evolution of ten EUV on-disk quiet-region  jets (between 2012 March and 2012 December), relating the jet-base field structure observed by \sdo/HMI to jet coronal components observed by  \sdo/AIA. Table \ref{tab:list} lists the ten jets and their measured parameters. In Section \ref{evo} we present the EUV observations of two jets from our list of ten jets. The pre-jet magnetic evolution is addressed in  Section \ref{flux}. 

\begin{figure*}
	\centering
	\includegraphics[width=\linewidth]{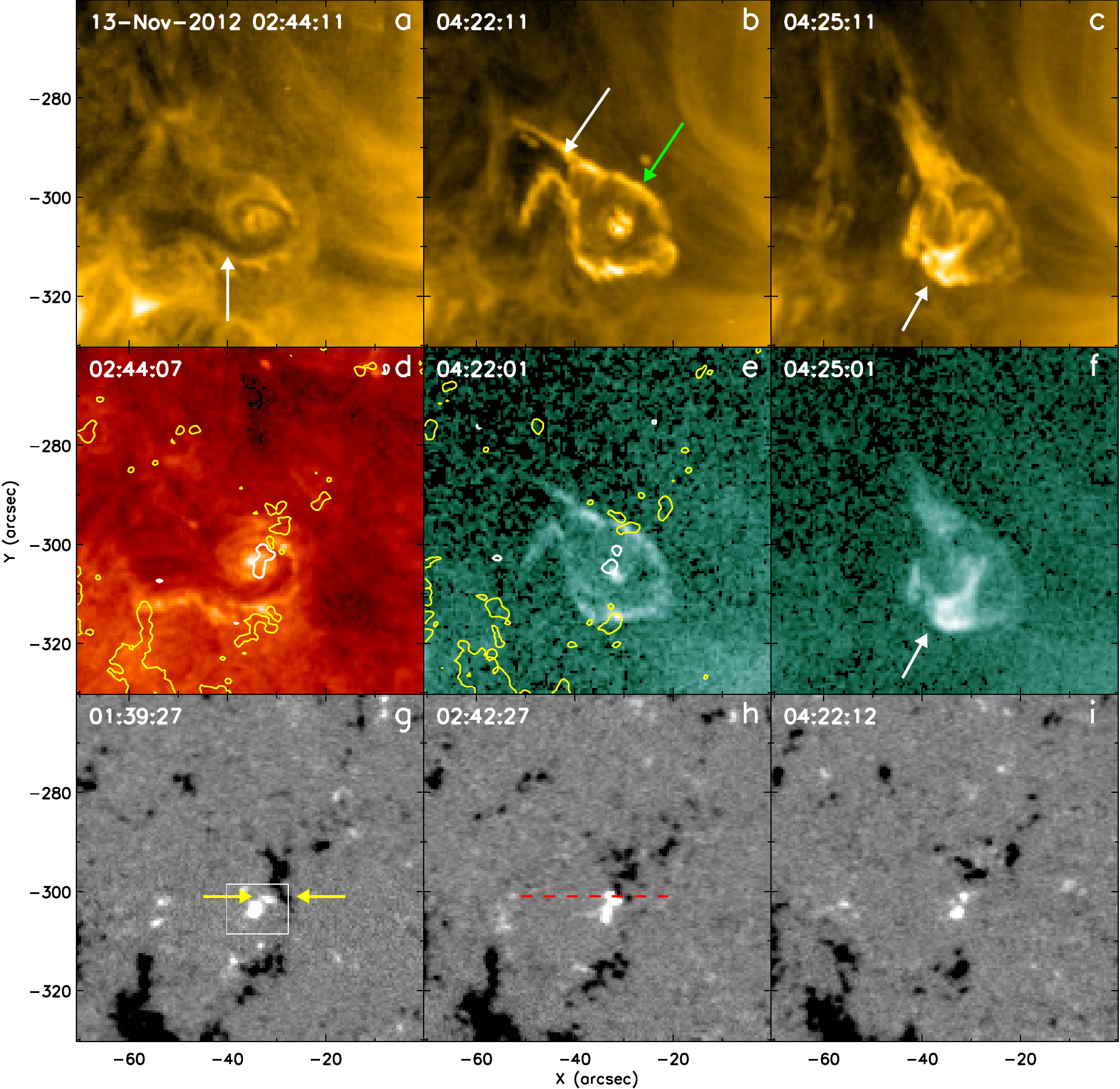} 
	\caption{Quiet region jet observed on 13-Nov-2012: (a-c) show 171 \AA, (d) 304 \AA\ and (e,f) 94 \AA\ AIA intensity images of the jet J9 of Table \ref{tab:list}; (g-i) show HMI magnetograms of the same region. In (a), the arrow shows the minifilament. In (b), the white arrow points to the minifilament segment that moves first during the jet eruption and the green arrow points to brightenings from external reconnection. In (c) and (f), the arrows point to the JBP. The yellow arrows in (g) point to positive and negative flux clumps converging on the neutral line on which the pre-jet minifilament resides; the box in (g) shows the region measured for the magnetic flux time plot in Figure \ref{fig3}c. The red dashed line in (h) shows the east-west cut for the time-distance image in Figure \ref{fig3}d. HMI contours (level $\pm$ 50 G) of (h) and (i) are overlaid on panels (d) and (e), respectively, where white and yellow respectively represent positive and negative
		polarities. Animations of this Figure ((a-c) and (g-i)) are available.} \label{fig2}
\end{figure*}

\begin{figure*}[ht]
	\centering
	\includegraphics[width=\linewidth]{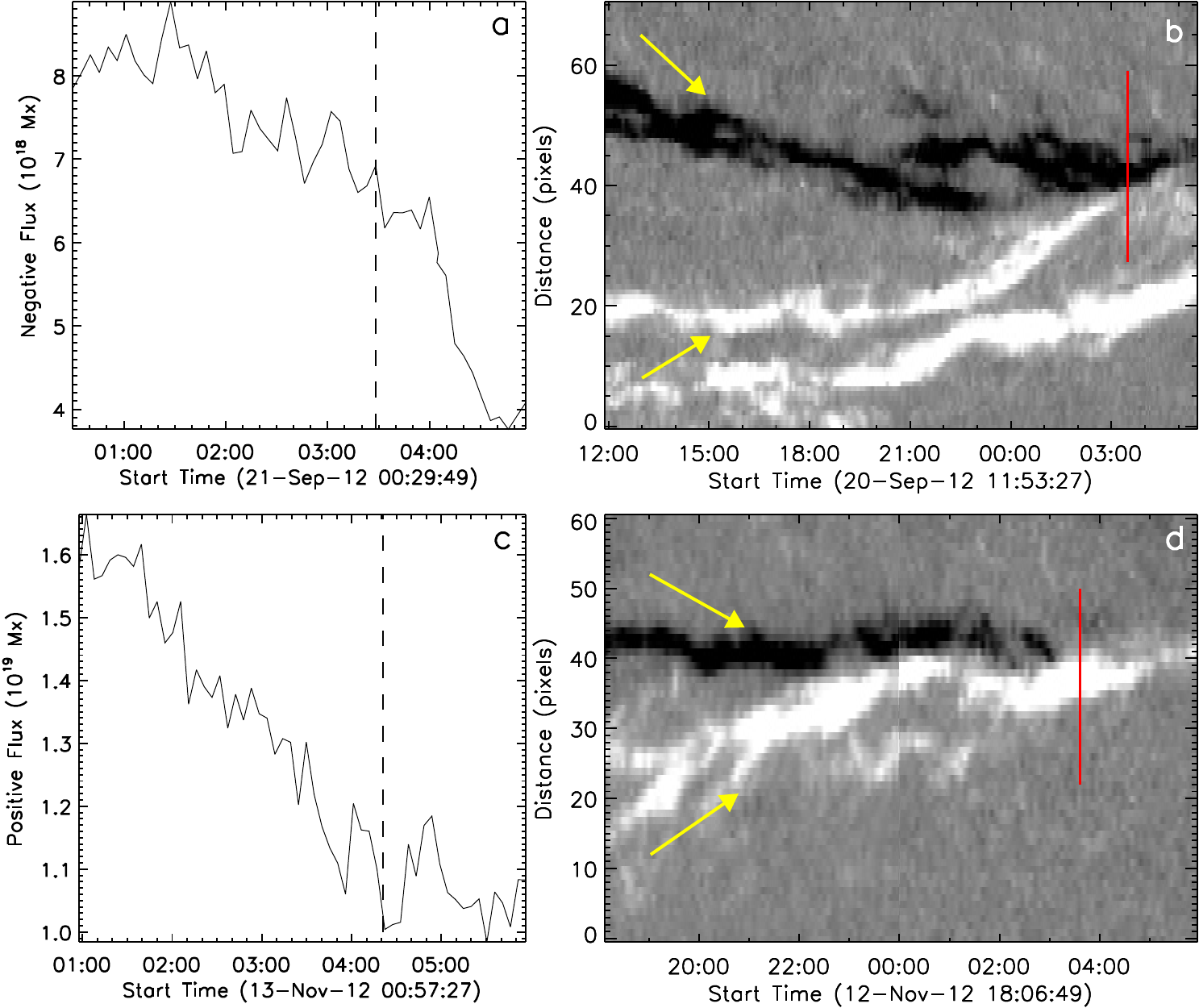}
	\caption{Jet-base magnetic flux evolution: (a) shows the negative flux as a function of time computed inside the box of Figure \ref{fig1}h. (c) shows the positive flux as a function of time measured inside the box region of Figure \ref{fig2}g. The dashed lines in (a) and (c) show the jet eruption time. Panels (b) and (d) show HMI  time-distance images along the red dashed lines in Figures \ref{fig1}h and \ref{fig2}h, respectively. The yellow arrows  in (b) and (d) point to the tracks of positive and negative flux clumps that converged and cancelled over $\sim$ 10 hours before each jet eruption (see Figures \ref{fig1} and \ref{fig2} magnetogram and their animations); the red lines in (b) and (d) show the jet eruption onset time.
	} \label{fig3} 
	
\end{figure*} 

\subsection{\textit{Evolution of Minifilaments and Jets}}\label{evo}

Figure \ref{fig1} shows a typical example of one of our EUV on-disk jets (J7, Table \ref{tab:list}). We observed a minifilament in the jet-base region (Figures \ref{fig1}a and d). It had a length of about 15000 km (Table \ref{tab:list}). Figure \ref{fig1}a shows the situation prior to the jet onset. Later, the minifilament takes part in the eruption. The video (MOVIE1) accompanying Figure \ref{fig1}(a-c), shows the evolution of the minifilament and jet. The miniflament starts to lift off at $\sim$ 03:25 UT and the JBP appears at 03:29 UT (see MOVIE1 and Figures \ref{fig1} (b), (c), (e), and (f)). The JBP occurs at the pre-eruption location of the minifilament. After the start of the JBP the jet spire starts to extend upwards, as shown in Figure \ref{fig1}(c) and (f). 

In Figure \ref{fig2}, we show another example jet (J9) from our list. The white arrow in Figure \ref{fig2}a points to a minifilament in the jet-base region. The minifilament has a length of $\sim$ 25000 km (Table \ref{tab:list}). 
Figure \ref{fig2}b shows the minifilament as it was rising slowly. The JBP starts to brighten at 04:23 UT (MOVIE2)  and later the spire becomes visible (Figure \ref{fig2}c). Figures \ref{fig2}(c) and (f) show that the JBP sits at the pre-eruption location of the minifilament. We do not observe any precursor brightenings at the location of the JBP in AIA 1600 and 1700 \AA\ images. In all ten events, the JBP brightening appears at the same time in the AIA EUV images and in the AIA 1600 \AA\ images. 

Similarly, the other eight jets resulted from eruption of minifilaments. Specifically, each of the jets resulted from the eruption of a minifilament from the site of the JBP\@.
All the jet-producing eruptions and JBPs studied here are similar to typical solar flare eruptions \citep[e.g.][]{mccauley15}, in which a flare arcade (analogous to the JBP) grows over the neutral line in the wake of the filament eruption. The minifilaments show a slow-rise, followed by a fast-rise as they erupt (MOVIE1 and MOVIE2), analogous to many longer-scale filament eruptions \citep[e.g.][]{sterling05,panesar15}.

We measured the length of the minifilaments, the jet-base widths, the jet speeds and jet durations (between the start and the maximum extent of the spire) from the AIA 171 \AA\ images. All of the measured parameters are given in Table \ref{tab:list}. We measured the plane-of-sky speeds of the jets by constructing a height-time plot for each of them.  For the uncertainties, we considered different fits to the height-time trajectories, and estimated the error in the speeds from the measured slopes. So for our two events above, we observed that J7 moves with a speed of 135 $\pm$ 30 \kms\ over about 12 minutes whereas J9 erupts outward with a speed of 55 $\pm$ 5 \kms\ over about 9 minutes.   

\subsection{\textit{Underlying Magnetic field}}\label{flux}

Figures \ref{fig1}(g-i) show  line-of-sight magnetograms before, during and after the eruption onset of jet J7. Figure \ref{fig1}d shows that the minifilament initially resides along a  neutral line in the quiet-region magnetic network. This is a region where positive flux is in the majority and the minifilament runs along the neutral line between majority and minority flux clumps (shown with yellow arrows in Figure \ref{fig1}g). We followed these positive and negative flux clumps before eruption and observed that they converged (see Figure \ref{fig1}(g-i) and MOVIE3). The yellow arrows in Figure \ref{fig1}(g) point to  positive and negative clumps that converge, merge and mostly cancel leading up to the minifilament eruption. 

To examine the evolution of the magnetic flux quantitatively we measured the negative flux of the jet region (J7), bounded by the white box of Figure \ref{fig1}h.
We measured only the negative flux because it is easy to isolate the patch of negative polarity. We  carefully checked that there were no negative-flux flows across the boundary of the box. Figure \ref{fig3}a shows the negative flux values, integrated over the selected region (white box of Figure \ref{fig1}h), as a function of time. The negative flux continuously decreases with time, which is clear evidence of flux cancellation at the neutral line of the minifilament. The dashed line in Figure \ref{fig3}a marks the eruption time, i.e. when we observe brightening in the EUV images (see MOVIE1) and the minifilament starts to lift off. To further display the flux convergence and cancellation, we created an HMI time-distance image ($\sim$ 18 hours, Figure \ref{fig3}b) along the red dashed line of Figure \ref{fig1}h. One can clearly see that both polarities approach the neutral line, and eventually cancel with each other just before the eruption (at 03:27 UT, red line in Figure \ref{fig3}b). The cancellation of the two clumps continues on even after the jet eruption (MOVIE3).

\begin{figure*}
	\centering
	\includegraphics[width=\linewidth]{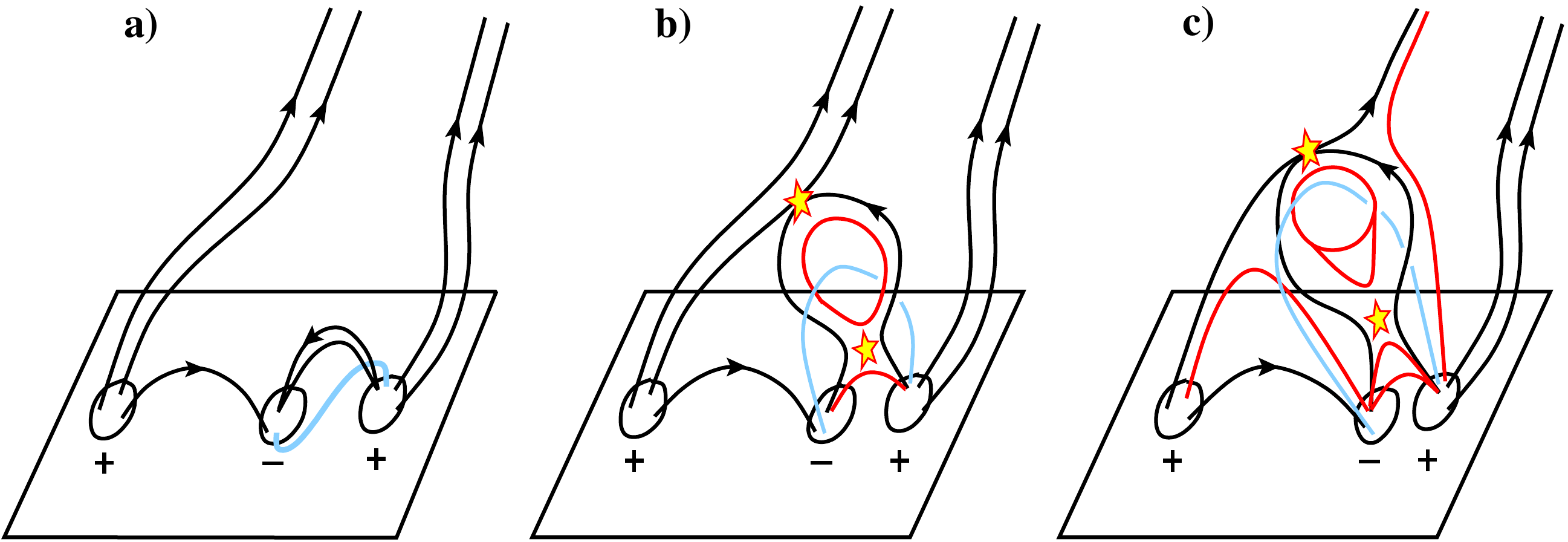}
	\caption{Schematic interpretation of the observations: The black lines and rectangular box represent the magnetic field and the solar surface, respectively. The blue line in (a), (b), and (c) represents sheared and twisted field holding a minifilament and the lines above it represent the field enveloping the minifilament field. Stars show locations where reconnection takes place. Red lines in (b) and (c) represent newly-reconnected field, which result from the internal (lower star)  and external (upper star) magnetic reconnection of the field enveloping the erupting twisted flux rope. The far-reaching red line in (c) shows the newly-reconnected open field along which the jet escapes.    
		The `+' sign labels positive polarity flux and the `--' sign is for  
		negative magnetic flux.} \label{fig4}
\end{figure*}
Figure \ref{fig2}(g-i) shows the magnetic flux arrangement for jet J9. The minifilament initially resides along the neutral line that runs between a clump of majority black flux and a clump of minority white flux at one end. 
Both clumps (shown with the yellow arrows in Figure \ref{fig2}g) merged and mostly cancelled with each other (see MOVIE4). For this jet we mainly focus on the changes in the positive (white) flux, which is the minority flux for this case, because it is more isolated:  its evolution is easy to follow and measure. The plot  in Figure \ref{fig3}c shows the evolution of the positive flux for $\sim$ 6 hours. The flux  continuously decreased through the time ($\sim$ 04:23 UT) of the appearance of the JBP at the pre-eruption site of the minifilament. This is clear evidence of flux cancellation at the location of minifilament leading to the eruption. The minifilament starts to lift off before the jet-spire onset. The time-distance image in Figure \ref{fig3}d shows the convergence and cancellation of the jet-base polarities (Figure \ref{fig1}h) for $\sim$ 12 hours. As the negative flux cancels and disappears, the jet eruption is  triggered at 03:50 UT.  

Similarly, we  tracked the evolution of magnetic flux in all eight other jet regions, and compared that evolution with the minifilament eruption onset time and find flux cancellation in each case. We estimated the percentage flux reduction from well before to after each of the ten events by calculating 
	the flux content of the minority polarity flux clumps 5-6 hours before and immediately (0-1 hours) after the jet eruption (Table \ref{tab:list}). We estimated the (1$\sigma$) uncertainty in flux decrease using images of the 5-min cadence flux values over the 1 hour windows (5-6 hours before and 0-1 hours after the eruption). These uncertainties are listed in Table \ref{tab:list}. 

We find that the triggering mechanism for all ten jet eruptions is evidently flux cancellation in the manner of our two example jets. This is consistent with the idea of other workers, e.g. \cite{hermans86}, \cite{jiang01}, \cite{sakajiri04}, and \cite{ren08}, that flux convergence and cancellation plays an important role in small-scale filament eruptions. In those listed studies, the small-scale filaments are smaller than typical solar filaments but larger than our minifilaments.
\vspace{0.3cm}
\section{SUMMARY AND DISCUSSION}\label{discussion}
\vspace{0.1cm}
We have examined in detail ten randomly-found on-disk quiet region jets observed by \sdo/AIA and \sdo/HMI. They appear to be typical coronal jets, and have these properties: (a) in each pre-jet jet base there was  cool transition region material, a minifilament, consistent with recent findings of \cite{sterling15}; (b) the jets shoot out from the solar surface with an average speed of 100 $\pm$ 20 \kms; (c) the average duration of the studied jets is 12 minutes; (d) the average jet-base width is 17000 $\pm$ 600 km; (e) the average length of the minifilament is 18 $\times$ 10$^{3}$ km, which is comparable to the jet-base width. In all cases, we observed a JBP at the neutral line from which the minifilament erupted. The jet-spire widths grew to about the width of the jet base; this is similar to what  \cite{moore10} found for their blowout jets, based on X-ray observations.

The observed jet speeds and durations are somewhat slower and shorter than those  obtained for quiet region X-ray jets by \cite{shimojo96} ($\sim$ 125 \kms, 65 minutes). Our jet durations are nearly the same as the average lifetime ($\sim$ 20 minutes) of EUV jets reported by \cite{nistico09}, and similar to that of the X-ray coronal hole jets of \cite{savcheva07} ($\sim$ 10 minutes). Our minifilament lengths are $\sim$ 2 times longer than those of \cite{sterling15} ($\sim$ 8 $\times$ 10$^{3}$ km); we measured  the total curvilinear length along the minifilament body for our quiet region minifilaments before eruption, while they measured the projected length of polar coronal hole minifilaments during eruption.


The minifilament eruptions are similar to many large filament eruptions, in showing slow-rise followed by fast-rise. The minifilaments initially lie on the magnetic neutral lines between the majority and minority flux and they erupt during flux cancellation at the neutral line. Magnetic flux cancellation is evidently the triggering mechanism in the studied jet eruptions (and evidently works in the way proposed by \cite{moore92} for larger filament eruptions). 

Figure \ref{fig4} shows a schematic sketch based on our jet observations. In Figure \ref{fig4}a, a smaller (explosive) bipole (on right hand side) is next a larger bipole (left hand side), where the smaller bipole contains a sheared and twisted field that holds a minifilament. The minifilament (blue) initially sits between the patches of majority positive flux and minority negative flux. The field immediately above (black loops in (a)) is an overlying pre-eruption arcade. Initially, the smaller bipole's footpoints are well separated from each other. In (b) and (c), we show that both polarities are approaching (and canceling)  each other (see Figures \ref{fig3}b,d). The resulting highly-sheared filament field eventually becomes unstable due to flux cancellation at the neutral line and erupts outward, resulting in internal reconnection in the erupting field (lower star in (b) and (c)). The JBP is the lower product of the internal reconnection, and is shown as low-lying (red) flare loop in (b) at the pre-eruption location of the minifilament. 
As it erupts, the outer envelope of the erupting minifilament field reconnects (known as external/interchange reconnection) with the  surrounding far-reaching coronal field above the large bipole (upper star in b and c; from AIA images we know that the surrounding-coronal field is often a far-reaching loop rather than truly open). The external reconnection results in two new connections: the red closed loop in (c) over the large bipole and the red open field line, which guides  along it the minifilament plasma that appears as part of the jet spire. We observe brightenings in EUV images  (Figures \ref{fig1}e and \ref{fig2}e) from the  external reconnection at the far ends of the red closed loops that are newly formed from the reconnection (see Figures \ref{fig1}b,  \ref{fig2}b and \ref{fig4}c). 

In summary, we report observations of ten random on-disk solar quiet region jets. We address the magnetic cause of the driving eruptions. Our observations show that in each case a  minifilament initially resides at a neutral line  inside the jet-base region, and flux cancellation at that neutral line triggers the jet-driving minifilament eruption.

\acknowledgments
N.K.P is supported by an appointment to the NASA Postdoctoral Program at the NASA MSFC, administrated by Universities Space Research Association under contract with NASA. This work was also funded by the Heliophysics Division
of NASA's Science Mission Directorate through the heliophysics Guest Investigators Program, and by the \Hinode\ Project. We acknowledge the use of the \sdo/AIA and \sdo/HMI
observations for this study. SDO data are courtesy of the NASA/\sdo\ AIA and HMI science teams. NKP thanks Dr. Anusha for useful discussion. We thank the referee for constructive comments.

\bibliographystyle{apj}

\end{document}